\documentstyle[12pt]{article}
\begin{document}
\setcounter{page}{0}
\thispagestyle{empty}
\begin{flushright}
 UTPT-94-35 \\
 hep-ph/9411412 \\
 November 1994 \\
\end{flushright}
\vspace{5mm}
\begin{center}
\Large
Momentum dependent quark mass in two-point correlators \\
\normalsize
\vspace{20mm}
B. Holdom and Randy Lewis \\
{\it Department of Physics \\
     University of Toronto \\
     Toronto, Ontario \\
     CANADA~~M5S~1A7} \\
\vspace{20mm}
ABSTRACT\\
\end{center}
A momentum dependent quark mass may be incorporated into a quark model in a
manner consistent with dynamically broken chiral symmetry.  We use this to
study the high $Q^2$ behavior of the vector, axialvector, scalar and
pseudoscalar two-point correlation functions. Expanding the
results to order $1/Q^6$, we show the correspondence between the dynamical
quark mass and the vacuum condensates which appear in the operator product
expansion of QCD.  We recover the correct leading logarithmic $Q^2$
dependence of the various terms in the OPE, but we also
find substantial subleading corrections which are numerically huge in a
specific case.  We conclude by using the vector minus axialvector correlator
to estimate the $\pi^+ - \pi^0$ electromagnetic mass difference.

\newpage
{\flushleft\large\bf Introduction}
\vspace{5mm}

For large momenta, the QCD running coupling is small and calculations may be
carried out systematically by a perturbative expansion in the coupling.
As the momentum scale is lowered, nonperturbative effects become significant
and the coupling expansion is no longer useful.  Consider specifically
the physics of light quark flavors in the context of a two-point
correlation function, $\Pi(q^2)$, which we define to be dimensionless.  From a
purely phenomenological standpoint, the onset of nonperturbative effects
can be parameterized by a series of correction terms.\cite{SVZ}
\begin{equation}\label{series}
   \tilde\Pi(q^2) = \Pi(q^2) - \Pi_{pert}(q^2) = \sum_{n=1}^N\frac{C_n}{Q^{2n}}
\end{equation}
Throughout this work, the tilde is used as a reminder that perturbative
physics has been excluded.
$Q^2=-q^2$ is the (Euclidean) momentum squared and the $C_n$ depend at most
logarithmically on $Q^2$.  When $Q^2$ (which we always choose to be positive)
is not too small, this series is dominated by the first few terms and can be
used to perform calculations at lower $Q^2$ than is possible using only the
perturbative coupling expansion.  (\ref{series}) is actually an operator
product expansion (OPE) and the $C_n$ are known functions of
various nonzero vacuum condensates.\cite{SVZ}

The chiral symmetry breaking vacuum condensate with the
lowest mass dimension is $\left<{\overline q}q\right>$, and it
(when multiplied by a current quark
mass) will contribute to $C_2$ in (\ref{series}).  There is one other
condensate in
$C_2$, $\left<\alpha_sG_{\mu\nu}^rG^{r\,\mu\nu}\right>$,
but it does not break chiral symmetry and will not be
of direct interest in what follows.  Since there are no condensates with mass
dimension two, $C_1$=0 in QCD.

The existence of a nonzero $\left<{\overline q}q\right>$
implies an effective quark
mass\cite{effmass}, and this in turn modifies the QCD Ward-Takahashi (WT)
identities for the couplings of vector and axialvector fields to quarks.
The complete two-point correlator in QCD is shown in Fig.~\ref{fig:loop},
where one vertex is a bare $\gamma_{\mu}[\gamma_5]$ but the other vertex
and both quark propagators represent fully-dressed nonperturbative quantities
that must satisfy the WT identities.
In what follows we will consider only the contributions of the effective
quark mass to these nonperturbative quantities,
and it will be shown that for both the vector and axialvector correlators,
 a ``minimal'' WT vertex
correctly reproduces $C_1$ and the $\left<{\overline q}q\right>$ term of $C_2$.
We obtain expressions for $C_3$ in the vector and axialvector cases
which have the known leading logarithmic running, and we point out the
possibility of substantial corrections in $C_3$ due to terms that are
formally suppressed by logarithms.

The same diagram (Fig.~\ref{fig:loop})
also represents the scalar and pseudoscalar two-point correlators in QCD,
although there are no WT identities to constrain the corresponding vertices.
If the full scalar or pseudoscalar vertex is approximated by a bare one,
$C_2$ is not generated in its correct form.
Improved scalar and pseudoscalar vertices
can be found by appealing to the gauged nonlocal constituent (GNC) quark
model.\cite{GNC}

The GNC Lagrangian contains constituent quarks with momentum dependent
masses and pseudoscalar mesons,
constructed to model dynamically broken chiral symmetry.  The couplings of
vector and axialvector fields to quarks are precisely
the ``minimal'' WT vertices mentioned
above.  Although the GNC Lagrangian originally included the scalar and
pseudoscalar fields in a trivial way\cite{GNC}, the Lagrangian allows for
a natural extension of these couplings in a manner analogous to its vector
and axialvector couplings.  For the case of the scalar
correlator we verify that this produces the correct expressions for the
chiral symmetry breaking pieces of $C_2$.  $C_3$ also has the correct leading
logarithmic running, but here we find enormous subleading correction
terms as well.

It is important to stress that we will use the GNC model only to
determine the form of the nonperturbative propagators and
vertices appearing in the general correlator of QCD in
Fig.~\ref{fig:loop}.  Our main goal is to study the relation between
the $1/Q^2$ expansion of these correlators and the momentum dependent mass
function.  We will need to consider the low energy behavior of these
correlators only when we treat the pion mass difference.  At low energies we
may apply the GNC model directly, since it has been shown to model low energy
phenomenology rather well\cite{GNC}\cite{moreGNC}.  Thus at low
energy the correlators will be described by the GNC model diagrams of
Fig.~\ref{fig:GNC}, and at high energies they will be described by
Fig.~\ref{fig:loop}.  The GNC model includes the pseudo-Goldstone bosons of
QCD, and virtual effects of these mesons are accounted for in the low
energy contribution according to the standard chiral Lagrangian approach.
Meson loops are naturally cutoff in the model at the point where these
particles lose their particle-like nature.

The vector minus axialvector two-point correlator is of special interest in
our study since it would vanish if chiral symmetry was not broken.  We will
choose an explicit form for the effective quark mass which becomes the known
form\cite{effmass} at large momentum scales and which resembles the
successful\cite{GNC}\cite{moreGNC}
GNC ansatz at small scales.  We can then calculate the vector minus axialvector
correlator numerically at any momentum scale by matching the low energy GNC
model (Fig.~\ref{fig:GNC} plus meson loops) to our high energy model
(Fig.~\ref{fig:loop}) at an intermediate scale.  An integral over all momenta
produces the $\pi^+ - \pi^0$ electromagnetic mass difference\cite{Das}.

\vspace{5mm}
{\flushleft\large\bf The general {\boldmath 1/$Q^2$} expansion}
\vspace{5mm}

It is convenient to express the
vector, axialvector, scalar and pseudoscalar two-point correlators in terms of
dimensionless functions of $Q^2$,
\begin{eqnarray}
   i{\int}d^4x~e^{iq{\cdot}x}\left<0|{\rm T}\{V_\mu^a(x)V_\nu^b(0)\}|0\right>
   & = & (q_{\mu}q_\nu-q^2g_{\mu\nu}){}^V\Pi^{ab}(Q^2) \\
   i{\int}d^4x~e^{iq{\cdot}x}\left<0|{\rm T}\{A_\mu^a(x)A_\nu^b(0)\}|0\right>
   & = & (q_{\mu}q_\nu-q^2g_{\mu\nu}){}^{A,1}\Pi^{ab}(Q^2)
              - q_{\mu}q_\nu{}^{A,0}\Pi^{ab}(Q^2) \\
   i{\int}d^4x~e^{iq{\cdot}x}\left<0|{\rm T}\{S^a(x)S^b(0)\}|0\right>
   & = & Q^2~{}^S\Pi^{ab}(Q^2) \\
   i{\int}d^4x~e^{iq{\cdot}x}\left<0|{\rm T}\{P^a(x)P^b(0)\}|0\right>
   & = & Q^2~{}^P\Pi^{ab}(Q^2)
\end{eqnarray}
where $a,b$ are SU($N_f$) flavor indices.  Throughout this work we
will restrict the discussion to light quarks.
According to Fig.~\ref{fig:loop} we must determine the full quark
propagator and the full vertex for each correlator.

The most general quark propagator is
\begin{equation}
   i{\cal S}(q) = \frac{iZ(Q^2)}{q\!\!\!/ - {\Sigma}(Q^2)}
\end{equation}
If we think of writing the nonperturbative contributions to the inverse
propagator as a $1/Q^2$ series, analogous to (\ref{series}), there are only
two condensates with mass dimension less than five:
$\left<{\overline q}q\right>$ and
$\left<\alpha_sG_{\mu\nu}^rG^{r\,\mu\nu}\right>$.
In the limit of vanishing current quark masses, we see on purely dimensional
grounds that $Z(Q^2)$ cannot contain $\left<{\overline q}q\right>$ while
${\Sigma}(Q^2)/Z(Q^2)$ is independent of
$\left<\alpha_sG_{\mu\nu}^rG^{r\,\mu\nu}\right>$.  For our analysis of
chiral symmetry breaking, we will set $Z(Q^2)$=1 at the expense of omitting
the gluon condensate (which contains no information about chiral symmetry
breaking) as well as other effects at higher order in the $1/Q^2$ series.
Recall that the $\pi^+ - \pi^0$ electromagnetic mass difference is independent
of $\left<\alpha_sG_{\mu\nu}^rG^{r\,\mu\nu}\right>$.

We will retain ${\Sigma}(Q^2)$ with the correct asymptotic
behavior\cite{effmass}
\begin{equation}\label{massform}
   {\Sigma}(Q^2) \rightarrow \frac{-4\pi\alpha_s(Q)}{3Q^2}
   \left<{\overline q}q\right>
   ~~~~,~{\rm as}~Q^2 \rightarrow \infty
\end{equation}
\begin{equation}\label{runfermion}
   \left<{\overline q}q\right> \equiv \left<{\overline q}q\right>_Q
        = \left<{\overline q}q\right>_\mu
          \left(\frac{\alpha_s(Q)}{\alpha_s(\mu)}\right)^{-d}
\end{equation}
\begin{equation}
   \alpha_s(Q) \rightarrow d\pi\left({\rm ln}\frac{Q^2}{\Lambda^2}\right)^{-1}
        ~~~~,~{\rm as}~Q^2 \rightarrow \infty
\end{equation}
$d=12/(33-2N_f)$ for $N_f$ quark flavors, $\mu$ is a renormalization
scale, $\Lambda$ is the QCD scale and the current quark masses are set to zero.
With this normalization, $\left<{\overline q}q\right>$
represents a single quark flavor summed
over three colors and summed over Dirac indices.

The vector and axialvector vertices ${}^{V,A}\Gamma_\mu^a(p,q)$ are
constrained to satisfy the Ward-Takahashi identities.
\begin{eqnarray}
   q^\mu{}^V\Gamma_\mu^a(p,p+q) & = & {\cal S}^{-1}(p+q)\frac{\lambda^a}{2} -
                                  \frac{\lambda^a}{2}{\cal S}^{-1}(p) \\
   q^\mu{}^A\Gamma_\mu^a(p,p+q) & = & {\cal S}^{-1}(p+q)\gamma_5\frac
                                  {\lambda^a}{2} + \gamma_5\frac{\lambda^a}
                                  {2}{\cal S}^{-1}(p)
\end{eqnarray}
The $\lambda^a$ are flavor generators normalized by
Tr$(\lambda^a\lambda^b)=2\delta^{ab}$.
These identities, plus the requirement that the vertices contain no
unphysical singularities, uniquely define the longitudinal part of the vector
and axialvector vertices.  We will choose the minimal vertices
by ignoring any extra transverse pieces that may exist.\cite{BallChiu}
The resulting vertices, for incoming(outgoing) quark momentum
$p$($p^{\prime}=p+q$),
are
\begin{eqnarray}\label{Vvertex}
   {}^V\Gamma_\mu^a(p,p^\prime) & = &
      \frac{\lambda^a}{2}\left[\gamma_\mu + (p+p^\prime)_\mu\left(\frac{
      {\Sigma}({P^\prime}^2)-{\Sigma}(P^2)}{{P^\prime}^2-P^2}\right)\right] \\
   {}^A\Gamma_\mu^a(p,p^\prime) & = & \label{Avertex}
      \frac{\lambda^a}{2}\left[\gamma_\mu - \frac{q_\mu}
      {q^2}\left({\Sigma}({P^\prime}^2)+{\Sigma}(P^2)\right)\right]\gamma_5
\end{eqnarray}
Again, $P^2({P^\prime}^2)=-p^2(-{p^\prime}^2)$.
Notice that the vector vertex is completely free of singularities (assuming
none are contained within ${\Sigma}(Q^2)$) but that the axialvector vertex is
required to have a singular point.  This massless state is the Goldstone
boson of the dynamically broken symmetry --- the pion for $N_f$=2.

There are no analogous identities to constrain the forms of the scalar and
pseudoscalar vertices.  One might be tempted to adopt the bare vertices
\begin{eqnarray}\label{simpleS}
   {}^S\Gamma^a(p,p+q) & = & -\frac{\lambda^a}{2} \\
   {}^P\Gamma^a(p,p+q) & = & i\frac{\lambda^a}{2}\gamma_5 \label{simpleP}
\end{eqnarray}
but this will lead to a disagreement with the OPE.  One must also decide how
to include current quark mass effects.  Our approach
will be to appeal to the GNC quark model which is the minimal Lagrangian
that contains constituent quarks with mass ${\Sigma}(Q^2)$ and which respects
the dynamically broken chiral symmetry.  The original GNC model\cite{GNC}
was used at low energy scales, and the external fields
were coupled to quarks according to (\ref{Vvertex}-\ref{simpleP}), but as
stated above, the scalar and pseudoscalar couplings contradict the OPE at
larger scales.
We therefore propose the following ``\,GNC$^\prime$\,'' model,
which is identical
to the original version except that the $S$ and $P$ fields now appear in
the path-ordered exponential as well as in the local term.  In Euclidean
spacetime,
\begin{eqnarray}\label{Lagrangian}
   {\cal L}_{\rm GNC^\prime}(x,y) & = &
      \overline{\psi}(x){\delta}(x-y)\gamma^\mu
      [\partial_\mu-iV_{\mu}(y)-i\gamma_5A_{\mu}(y)+\frac{1}{4}\gamma_{\mu}S(y)
      +\frac{i}{4}\gamma_\mu\gamma_5P(y)]{\psi}(y) \nonumber \\
   &  & +\overline{\psi}(x){\Sigma}(x-y){\xi}(x)X(x,y){\xi}(y){\psi}(y) \\
   X(x,y) & = & {\rm P}\exp\left[-i\int_x^y\Gamma_{\mu}(z)dz^{\mu}\right] \\
   \Gamma_\mu(z) & = & \frac{i}{2}\xi(z)[\partial_\mu-iV_\mu(z)-i\gamma_5
      A_{\mu}(z)-\frac{1}{2}\gamma_{\mu}S(z)-\frac{i}{2}\gamma_\mu\gamma_5
      P(z)]\xi^\dagger(z) + \nonumber \\
  &&  \frac{i}{2}\xi^\dagger(z)[\partial_\mu-iV_{\mu}(z)+i\gamma_5A_{\mu}(z)-
      \frac{1}{2}\gamma_{\mu}S(z)+\frac{i}{2}\gamma_\mu\gamma_5P(z)]\xi(z) \\
   {\xi}(x) & = & \exp\left[\frac{-i\gamma_5}{f}\sum_{a=1}^8
                          \frac{\lambda^a}{2}{\pi^a}(x)\right]
\end{eqnarray}
$\pi^a$ contains the $N_f^2$-1 pseudoscalar mesons with decay constant $f$, and
$\psi$ is the $N_f$-plet of quark fields with mass ${\Sigma}(Q^2)$, the
Fourier transform
of ${\Sigma}(x-y)$.  $V_\mu$, $A_\mu$, $S$, $P$ are the external fields.
$X(x,y)$ is a path-ordered exponential.  Notice that we
have assumed the same mass function ${\Sigma}(Q^2)$ for all quark flavors,
which means that $\left<{\overline q}q\right>$ is also flavor-independent.

The Feynman rules for the GNC$^\prime$ model can be obtained from the
Lagrangian (\ref{Lagrangian}) in a systematic manner as described in
\cite{Terning}.  Notice
that the meson pole which contributes to the full axialvector and pseudoscalar
vertices now appears explicitly, as shown in Fig.~\ref{fig:pole}.  Although the
GNC$^\prime$ Lagrangian does not contain an explicit meson propagator, a
propagator is generated by the quark loops of Fig.~\ref{fig:prop}.  The
explicit calculation shows that the GNC$^\prime$ Lagrangian does indeed
reproduce the vector and axialvector vertices of
(\ref{Vvertex}-\ref{Avertex}).
To include the effects of a nonzero current quark mass matrix, $\cal M$, we
make the substitution
\begin{equation}
   S(x) \rightarrow {\cal M} + S(x)
\end{equation}
For our purposes, we will only require terms linear in $\cal M$.

With Feynman rules in hand, the expressions for the two-point correlators
of Fig.~\ref{fig:loop} are
easily written down in the form of 4-momentum integrals.
The integrands are largest when the momentum-squared flowing through one
quark propagator is of order ${\Sigma}^2(0)$ so that the other propagator has
a momentum-squared of order $Q^2$.
When the results are expanded in powers of $\cal M$ and $1/Q^2$,
we obtain a series of the form (\ref{series}).  The simple procedure
of expanding the {\it integrand\/} in $1/Q^2$ generates integral expressions
for the various $C_n$, and these integrals become more divergent for
increasing $n$. We stress that the full expression for each $\tilde\Pi(Q^2)$
is finite and it is only the simple expansion technique which creates
apparent divergences. Our results are
\begin{eqnarray}\label{Vexpand}
   {}^{V}\tilde\Pi^{ab}(Q^2) & = &
    \frac{{\cal M}^{ab}}{4Q^4}(-\Upsilon_1+\ldots)
    - \frac{\delta^{ab}}{Q^6}\left(
    \Upsilon_1Q^2{\Sigma}(Q^2)+\frac{\Upsilon_2}{3}+\ldots\right)
    + {\cal O}\left(\frac{1}{Q^8}\right) \\
   {}^{A,1}\tilde\Pi^{ab}(Q^2) & = &
    \frac{{\cal M}^{ab}}{4Q^4}(\Upsilon_1+\ldots)
    + \frac{\delta^{ab}}
    {Q^6}\left(\Upsilon_1Q^2{\Sigma}(Q^2)+\frac{\Upsilon_2}{6}+\ldots\right)
    + {\cal O}\left(\frac{1}{Q^8}\right) \label{A1expand} \\
   {}^{A,0}\tilde\Pi^{ab}(Q^2) & = &
    \frac{{\cal M}^{ab}}{2Q^4}(-\Upsilon_1+\ldots)
    + {\cal O}\left(\frac{1}{Q^8}\right) \\
   {}^{S}\tilde\Pi^{ab}(Q^2) & = &
    \frac{{\cal M}^{ab}}{8Q^4}(-3\Upsilon_1+\ldots)
    - \frac{\delta^{ab}}{Q^6}\left(\frac{3\Upsilon_1}{2}Q^2{\Sigma}(Q^2)
    - \Upsilon_2 + \ldots \right)
    + {\cal O}\left(\frac{1}{Q^8}\right) \label{Sexpand} \\
   {}^{P}\tilde\Pi^{ab}(Q^2) & = & {\cal O}\left(\frac{1}{Q^2}\right)
      \label{Pexpand}
\end{eqnarray}
We have defined
\begin{equation}
   {\cal M}^{ab} =
   {\rm Tr}\left({\cal M}\left\{\lambda^a,\lambda^b\right\}\right)
\end{equation}
where $\left\{\lambda^a,\lambda^b\right\}$ is an anticommutator.  The
ellipses at a given order in $1/Q^2$ denote the undetermined contributions
from divergent integrals which exist at higher order in the naive integrand
expansion.  (We will discuss the calculation of these terms below.) We have
not shown the current mass matrix $\cal M$ except for the leading
logarithmic terms at order $Q^{-4}({\rm ln}Q^2)^{d}$, where we have
retained the terms linear in $\cal M$. The $\Upsilon_n$ are one-dimensional
integrals which diverge logarithmically and require renormalization.
When (\ref{massform}) is used to evaluate these
integrals from the QCD scale $\Lambda$ up to a cutoff $Q^2$, we obtain
\begin{eqnarray}\label{ups1}
   \Upsilon_1 & \equiv & \frac{3}{4\pi^2}\int_{\Lambda^2}^{Q^2}~dx\frac
                {x{\Sigma}(x)}{x+\Sigma^2(x)} = -\left<{\overline q}q\right> \\
   \Upsilon_2 & \equiv & \frac{3}{4\pi^2}\int_{\Lambda^2}^{Q^2}~dx\frac
                {x^2\Sigma^2(x)}{x+\Sigma^2(x)} =
                \frac{-4d^2}{3(1-2d)}\pi^2\left<{\overline q}q\right>^2
                \left({\rm ln}\,\frac{Q^2}{\Lambda^2}\right)^{-1} \label{ups2}
\end{eqnarray}

To leading order in $\alpha_s$, the OPE in QCD gives\cite{SVZ}
\begin{eqnarray}
   {}^{V}\tilde\Pi^{ab}(Q^2) & = & \frac{{\cal M}^{ab}}{4Q^4}
    \left<{\overline q}q\right>
    + \frac{\delta^{ab}}{24{\pi}Q^4}
      \left<\alpha_sG_{\mu\nu}^rG^{r\,\mu\nu}\right>
       - \frac{{\cal A}^{ab}(\gamma_\mu\gamma_5)}{2Q^6}
    - \frac{{\cal B}^{ab}}{18Q^6}
    + {\cal O}\left(\frac{1}{Q^8}\right) \label{Vcond} \\
   {}^{A,1}\tilde\Pi^{ab}(Q^2) & = & \frac{-{\cal M}^{ab}}{4Q^4}
      \left<{\overline q}q\right>
    + \frac{\delta^{ab}}{24{\pi}Q^4}
      \left<\alpha_sG_{\mu\nu}^rG^{r\,\mu\nu}\right>
       - \frac{{\cal A}^{ab}(\gamma_\mu)}{2Q^6}
    - \frac{{\cal B}^{ab}}{18Q^6}
       + {\cal O}\left(\frac{1}{Q^8}\right) \label{A1cond} \\
   {}^{A,0}\tilde\Pi^{ab}(Q^2) & = & \frac{-{\cal M}^{ab}}{2Q^4}
         \left<{\overline q}q\right>
       + {\cal O}\left(\frac{1}{Q^8}\right) \label{A0cond} \\
   {}^{S}\tilde\Pi^{ab}(Q^2) & = & \frac{3{\cal M}^{ab}}{8Q^4}
      \left<{\overline q}q\right>
    + \frac{\delta^{ab}}{16{\pi}Q^4}
      \left<\alpha_sG_{\mu\nu}^rG^{r\,\mu\nu}\right>
       + \frac{\sqrt{4\pi\alpha_s}}{16Q^6}{\cal M}^{ab}
     \left<{\overline q}\sigma_{\mu\nu}\tau^rqG^{r\,\mu\nu}\right> \nonumber \\
    && + \frac{{\cal A}^{ab}(\sigma_{\mu\nu})}{4Q^6}
       + \frac{{\cal B}^{ab}}{12Q^6}
       + {\cal O}\left(\frac{1}{Q^8}\right) \label{Scond} \\
   {}^{P}\tilde\Pi^{ab}(Q^2) & = & \frac{-{\cal M}^{ab}}{8Q^4}
      \left<{\overline q}q\right>
    + \frac{\delta^{ab}}{16{\pi}Q^4}
      \left<\alpha_sG_{\mu\nu}^rG^{r\,\mu\nu}\right>
       + \frac{{\cal A}^{ab}(\sigma_{\mu\nu}\gamma_5)}{4Q^6}
       + \frac{{\cal B}^{ab}}{12Q^6}
       + {\cal O}\left(\frac{1}{Q^8}\right)~~~~ \label{Pcond}
\end{eqnarray}
We have defined
\begin{eqnarray}
   {\cal A}^{ab}(\Gamma_{\mu\ldots}) & = & \pi\alpha_s
             \left<{\overline\psi}\Gamma_{\mu\ldots}\tau^r\lambda^a\psi
             {\overline\psi}\Gamma^{\mu\ldots}\tau^r\lambda^b\psi\right> \\
   {\cal B}^{ab} & = & \pi\alpha_s\left<{\overline\psi}\gamma_\mu\tau^r
             \left\{\lambda^a,\lambda^b\right\}\psi\sum_{q=u,d,s,\ldots}
             {\overline q}\gamma^\mu\lambda^rq\right>
\end{eqnarray}
Recall that $q$ denotes a single quark flavor whereas $\psi$ is the quark
$N_f$-plet.  The $\tau^r$ are SU(3) color generators normalized
by Tr$(\tau^r\tau^s)=2\delta^{rs}$.

A comparison of (\ref{Vexpand}-\ref{Pexpand}) and (\ref{Vcond}-\ref{Pcond})
gives our expressions for the leading vacuum condensates.
At the outset of this section we neglected
$\left<\alpha_sG_{\mu\nu}^rG^{r\,\mu\nu}\right>$ contributions to the
quark propagator, and we now see explicitly that
$\left<\alpha_sG_{\mu\nu}^rG^{r\,\mu\nu}\right>$ is absent from
our results for the correlators.
The results verify, for all correlators except possibly the
pseudoscalar, that the model correctly generates no dimension two
condensates.  The integrand
of the pseudoscalar expression could not be expanded in our simple manner
beyond dimension two without introducing quadratically divergent integrals.
Whether or not the offending terms actually sum to zero cannot be determined
with this technique. At order $1/Q^4$, the terms in the vector, axialvector
and scalar correlators which involve the current quark masses also agree with
the OPE if we can neglect the ellipses in (\ref{Vexpand}-\ref{Sexpand}).
(We will justify the neglect of these terms below.)  The successful
result for the scalar correlator in particular is a reflection of the scalar
vertex present in the GNC$^\prime$ Lagrangian.

At dimension six,
we neglect $\cal M$ corrections and find expressions for the 4-quark
condensates that appear in each of three correlators.
\begin{eqnarray}
   9{\cal A}^{ab}(\gamma_\mu) + {\cal B}^{ab} & = &
   -18\Upsilon_1Q^2\Sigma(Q^2)\delta^{ab}-3\Upsilon_2\delta^{ab} + \ldots
   \nonumber \\
   & = & -4d\left(6-\frac{d}{(1-2d)}\right)\pi^2\delta^{ab}
   \left<{\overline q}q\right>^2
   \left({\rm ln}\,\frac{Q^2}{\Lambda^2}\right)^{-1} + \ldots
   \label{defcond1} \\
   9{\cal A}^{ab}(\gamma_\mu\gamma_5) + {\cal B}^{ab} & = &
   18\Upsilon_1Q^2\Sigma(Q^2)\delta^{ab}+6\Upsilon_2\delta^{ab} + \ldots
   \nonumber \\
   & = & 4d\left(6-\frac{2d}{(1-2d)}\right)\pi^2\delta^{ab}
   \left<{\overline q}q\right>^2
   \left({\rm ln}\,\frac{Q^2}{\Lambda^2}\right)^{-1} + \ldots
   \label{defcond2} \\
   3{\cal A}^{ab}(\sigma_{\mu\nu}) + {\cal B}^{ab} & = &
   -18\Upsilon_1Q^2\Sigma(Q^2)\delta^{ab}+12\Upsilon_2\delta^{ab} + \ldots
   \nonumber \\
   & = & -4d\left(6+\frac{4d}{(1-2d)}\right)\pi^2\delta^{ab}
   \left<{\overline q}q\right>^2
   \left({\rm ln}\,\frac{Q^2}{\Lambda^2}\right)^{-1} + \ldots \label{defcond3}
\end{eqnarray}
Recollection of the $Q^2$ dependence of $\left<{\overline q}q\right>$
given in (\ref{runfermion})
reveals that these functions exhibit the known $({\rm ln}Q^2)^{2d-1}$
dependence on $Q^2$.\cite{SVZ}  However, the presence of undetermined
contributions at the same order in $1/Q^2$
prevents us from using this approach to obtain complete
expressions for these condensates.  To proceed, we examine a specific example
in detail.

\vspace{5mm}
{\flushleft\large\bf Rigorous {\boldmath 1/$Q^2$} expansion for a specific
                     mass function}
\vspace{5mm}

The loop integrations from Fig.~\ref{fig:loop} can be performed rigorously for
the simple mass function which has typically been used in low energy GNC
calculations.
\begin{equation}\label{oldmass}
   \Sigma(Q^2) = \frac{(A+1)m_0^3}{Am_0^2+Q^2}
\end{equation}
$m_0 \sim 300$MeV represents the scale of the constituent quark mass and the
data require
$2 \,\,\lower 2pt \hbox{$\buildrel<\over{\scriptstyle{\sim}}$}\,\, A
\,\,\lower 2pt \hbox{$\buildrel<\over{\scriptstyle{\sim}}$}\,\, 3$.\cite{GNC}
Our goal is to build a mass function which becomes
(\ref{oldmass})[(\ref{massform})] in the limit of small[large] $Q^2$, but we
begin with a discussion of (\ref{oldmass}) alone.

For any $A>1.44$, the Euclidean
propagator can be expanded in a convergent geometric series.
\begin{eqnarray}
   \frac{1}{Q^2+\Sigma^2(Q^2)} & = & \frac{1}{Q^2+Am_0^2}\sum_{k=0}^\infty
   \left(\frac{Am_0^2-\Sigma^2(Q^2)}{Q^2+Am_0^2}\right)^k \\
   & = & \sum_{k=0}^\infty\sum_{l=0}^k\frac{k!}{l!(k-l)!}
   \frac{(Am_0^2)^{k-l}(-(A+1)^2m_0^6)^l}{(Q^2+Am_0^2)^{k+2l+1}}
\end{eqnarray}
When this relation is used for each propagator in the expression for a
two-point correlator, the
result contains four summations, but each term in the infinite series
is an integral that
can be evaluated by introducing a Feynman parameter in the standard fashion.
The final expression does not, in general, resum to a simple functional form,
but can be reduced to a single summation over hypergeometric functions which
can then be summed numerically to any desired accuracy.

For example, the leading nonperturbative behavior of the vector correlator is
\begin{equation}
   {}^V\tilde\Pi^{ab}(Q^2) = \frac{-12}{(4\pi)^2}{\cal M}^{ab}tu^2
    \left[{\rm ln}\left(\frac{1}{u}\right)-Y_V(v)\right] + {\cal O}(u^3)
\end{equation}
where $Y_V(v)$ denotes the infinite series and
\begin{eqnarray}
   t & \equiv & \frac{A+1}{A^2m_0} \rightarrow \frac{\Sigma^2(0)}
       {Q^2\Sigma(Q^2)}~~~{\rm as~} Q^2 \rightarrow \infty \\
   u & \equiv & \frac{Am_0^2}{Q^2} \rightarrow \frac{\Sigma(Q^2)}{\Sigma(0)}
       ~~~{\rm as~} Q^2 \rightarrow \infty \label{defu} \\
   v & \equiv & \frac{(A+1)^2}{A^3} \rightarrow \frac{\Sigma^3(0)}
       {Q^2\Sigma(Q^2)}~~~{\rm as~} Q^2 \rightarrow \infty \label{defv}
\end{eqnarray}
For any $A>1.44$, $Y_V(v)$ is well-approximated by the following simple form.
\begin{equation}
   Y_V(v) \approx \frac{1}{2} - 0.45v^{0.44}
\end{equation}
The value of $Y_V(0)$ is exact, and all derivatives at $v$=0 are infinite in
the true (series) expression as well as in this approximation.

If we assume (incorrectly) that this result remains valid when (\ref{oldmass})
is modified to be consistent with the correct asymptotic form of
(\ref{massform}), then we get
\begin{eqnarray}
    {}^V\tilde\Pi^{ab}(Q^2) & = & \frac{d}{4Q^4}{\cal M}^{ab}
    \left<{\overline q}q\right>\left(
    {\rm ln}\frac{Q^2}{\Lambda^2}\right)^{-1}
    \left[{\rm ln}\left(\frac{\tilde{v}Q^2}{\Sigma^2(0)}\right)
    -Y_V(\tilde{v})\right] + {\cal O}(u^3)\label{aa} \\
    \tilde{v} & \equiv & \frac{\Sigma^3(0)}{Q^2\Sigma(Q^2)} =
                \frac{-3\Sigma^3(0)}{4d\pi^2\left<{\overline q}q\right>}
                \left({\rm ln}\frac{Q^2}{\Lambda^2}\right)
\end{eqnarray}
A comparison with (\ref{Vcond}) shows that we have obtained the correct
leading logarithm except for a missing factor of $1/d$.  This is to be
expected.
If we had put the true form of the mass function (\ref{massform}) into the
integral of Fig.~\ref{fig:loop}, there would have been an extra factor of $1/d$
which arises from the integration.  This is easily demonstrated by computing
the large $Q^2$ behavior of $\Upsilon_1$, as defined in (\ref{ups1}), with a
cutoff.
\begin{equation}\label{extrad}
   \int_{\Lambda^2}^{Q^2}dx\frac{x\Sigma(x)}{x} \sim
   \int_{\Lambda^2}^{Q^2}dx\frac{1}{x}\left({\rm ln}\frac{x}{\Lambda^2}
   \right)^{d-1} \sim \frac{1}{d}\left({\rm ln}\frac{Q^2}{\Lambda^2}\right)^d
\end{equation}
It is now clear that we were justified in neglecting the uncomputed terms
containing $\cal M$ in (\ref{Vexpand}-\ref{Sexpand}) as was postulated in the
previous section --- the uncomputed terms produce
functions $Y_V(v)$, $Y_A(v)$ and $Y_S(v)$ which are logarithmically
suppressed relative to the computed term.

We now set $\cal M$=0 and consider the leading nonperturbative contributions to
the vector, axialvector and scalar correlators.  The pseudoscalar correlator
could also be obtained, but the presence of the pion propagator
(Figs.~\ref{fig:pole} and \ref{fig:prop}) makes the calculation rather tedious.
The pion contribution to the axialvector correlator is
slightly more pleasant, and the transverse nature of this
correlator provides a valuable check on the calculation.
The relation between the vector minus axialvector correlator and the pion mass
difference, which we discuss below, is another incentive for pursuing the
axialvector calculation.  Our results are
\begin{eqnarray}
   {}^V\tilde\Pi^{ab}(Q^2) & = & \frac{34\delta^{ab}}{(4\pi)^2}u^3v
    \left[{\rm ln}\left(\frac{1}{u}\right)-Z_V(v)\right] + {\cal O}(u^4) \\
   {}^{A,1}\tilde\Pi^{ab}(Q^2) & = & \frac{-14\delta^{ab}}{(4\pi)^2}u^3v
    \left[{\rm ln}\left(\frac{1}{u}\right)-Z_A(v)\right] + {\cal O}(u^4) \\
   {}^S\tilde\Pi^{ab}(Q^2) & = & \frac{-15\delta^{ab}}{(4\pi)^2}u^3v
    \left[{\rm ln}\left(\frac{1}{u}\right)-Z_S(v)\right] + {\cal O}(u^4)
\end{eqnarray}
where $Z_X(v)$ denote the infinite series and $u$, $v$ are defined in
(\ref{defu}-\ref{defv}).  For any $A>1.44$,
the $Z_X(v)$ are well-approximated by the following simple forms.
\begin{eqnarray}\label{ZV}
   Z_V(v) & \approx & \frac{67}{51} + 0.24\sqrt{v} \\
   Z_A(v) & \approx & \frac{1}{42} + 0.40\sqrt{v} \\
   Z_S(v) & \approx & \frac{67}{5} + \frac{1.20}{\sqrt{v}}{\rm ln}(1+v)
            \label{ZS}
\end{eqnarray}
The values of $Z_X(0)$ are exact, and all derivatives at $v$=0 are infinite in
the true (series) expressions as well as in these approximations.

If we consider a modification of (\ref{oldmass}) to make it consistent with
the correct asymptotic form of (\ref{massform}), then
we see that each correlator regains the known leading logarithmic behavior of
(\ref{defcond1}-\ref{defcond3}).  Moreover, we can now estimate the size of
these condensates.
\begin{eqnarray}\label{estimateV}
   9{\cal A}^{ab}(\gamma_\mu) + {\cal B}^{ab} & = &
      28d^2\pi^2\delta^{ab}\left<{\overline q}q\right>^2\left({\rm
ln}\frac{Q^2}
      {\Lambda^2}\right)^{-2}\left[{\rm ln}\left(\frac{\tilde{v}Q^2}
      {\Sigma^2(0)}\right)-Z_A(\tilde{v})\right] \\
   9{\cal A}^{ab}(\gamma_\mu\gamma_5) + {\cal B}^{ab} & = &
     -68d^2\pi^2\delta^{ab}\left<{\overline q}q\right>^2\left({\rm
ln}\frac{Q^2}
      {\Lambda^2}\right)^{-2}\left[{\rm ln}\left(\frac{\tilde{v}Q^2}
      {\Sigma^2(0)}\right)-Z_V(\tilde{v})\right] \\
   3{\cal A}^{ab}(\sigma_{\mu\nu}) + {\cal B}^{ab} & = &
     -20d^2\pi^2\delta^{ab}\left<{\overline q}q\right>^2\left({\rm
ln}\frac{Q^2}
      {\Lambda^2}\right)^{-2}\left[{\rm ln}\left(\frac{\tilde{v}Q^2}
      {\Sigma^2(0)}\right)-Z_S(\tilde{v})\right] \label{estimateS}
\end{eqnarray}
We may compare these to those in the vacuum saturation approximation:
\begin{eqnarray}
   9{\cal A}^{ab}(\gamma_\mu) + {\cal B}^{ab} & = &
      \frac{224}{9}d\pi^2\delta^{ab}\left<{\overline q}q\right>^2
      \left({\rm ln}\frac{Q^2}{\Lambda^2}\right)^{-1} \\
   9{\cal A}^{ab}(\gamma_\mu\gamma_5) + {\cal B}^{ab} & = &
      \frac{-352}{9}d\pi^2\delta^{ab}\left<{\overline q}q\right>^2
      \left({\rm ln}\frac{Q^2}{\Lambda^2}\right)^{-1} \\
   3{\cal A}^{ab}(\sigma_{\mu\nu}) + {\cal B}^{ab} & = &
      \frac{-352}{9}d\pi^2\delta^{ab}\left<{\overline q}q\right>^2
      \left({\rm ln}\frac{Q^2}{\Lambda^2}\right)^{-1}
\end{eqnarray}
The two expressions have the same leading logarithmic dependence on $Q^2$,
but the numerical factors in front differ.  We could make similar remarks
here as made in the discussion surrounding (\ref{aa}-\ref{extrad}).

Of greater interest are the terms which are subleading to the leading
logarithm, and which correspond to terms which are often neglected in
approximations to QCD such as vacuum saturation.  We may compare the terms
in brackets in (\ref{aa}) and (\ref{estimateV}-\ref{estimateS}) to the
leading logarithm, ln($Q^2/\Lambda^2$).  By making reasonable estimates of
the  various quantities involved,
\begin{eqnarray}
  & 100{\rm MeV} \,\,\lower 2pt
\hbox{$\buildrel<\over{\scriptstyle{\sim}}$}\,\,
    \Lambda \,\,\lower 2pt \hbox{$\buildrel<\over{\scriptstyle{\sim}}$}\,\,
    300{\rm MeV} \\
  & 400{\rm MeV} \,\,\lower 2pt
\hbox{$\buildrel<\over{\scriptstyle{\sim}}$}\,\,
    \Sigma(0) \,\,\lower 2pt \hbox{$\buildrel<\over{\scriptstyle{\sim}}$}\,\,
    500{\rm MeV} \\
   & -(230{\rm MeV})^3
    \,\,\lower 2pt \hbox{$\buildrel<\over{\scriptstyle{\sim}}$}\,\,
    \left<{\overline q}q\right>_\mu\left({\rm ln}\frac{\mu^2}{\Lambda^2}
    \right)^{-d} \,\,\lower 2pt
\hbox{$\buildrel<\over{\scriptstyle{\sim}}$}\,\,
    -(180{\rm MeV})^3
\end{eqnarray}
and by setting $d$=4/9 and $Q$=1GeV we obtain
\begin{eqnarray}
   2.4 \,\,\lower 2pt \hbox{$\buildrel<\over{\scriptstyle{\sim}}$}\,\, &
   {\rm ln}\left(\frac{Q^2}{\Lambda^2}\right) &
   \,\,\lower 2pt \hbox{$\buildrel<\over{\scriptstyle{\sim}}$}\,\, 4.6 \\
   2.2 \,\,\lower 2pt \hbox{$\buildrel<\over{\scriptstyle{\sim}}$}\,\, &
   {\rm ln}\left(\frac{\tilde{v}Q^2}{\Sigma^2(0)}\right)
               -Y_V(\tilde{v}) &
   \,\,\lower 2pt \hbox{$\buildrel<\over{\scriptstyle{\sim}}$}\,\, 4.2 \\
   1.7 \,\,\lower 2pt \hbox{$\buildrel<\over{\scriptstyle{\sim}}$}\,\, &
   {\rm ln}\left(\frac{\tilde{v}Q^2}{\Sigma^2(0)}\right)
               -Z_A(\tilde{v}) &
   \,\,\lower 2pt \hbox{$\buildrel<\over{\scriptstyle{\sim}}$}\,\, 2.3 \\
   0.6 \,\,\lower 2pt \hbox{$\buildrel<\over{\scriptstyle{\sim}}$}\,\, &
   {\rm ln}\left(\frac{\tilde{v}Q^2}{\Sigma^2(0)}\right)
               -Z_V(\tilde{v}) &
   \,\,\lower 2pt \hbox{$\buildrel<\over{\scriptstyle{\sim}}$}\,\, 1.5 \\
   -12.1 \,\,\lower 2pt \hbox{$\buildrel<\over{\scriptstyle{\sim}}$}\,\, &
   {\rm ln}\left(\frac{\tilde{v}Q^2}{\Sigma^2(0)}\right)
                 -Z_S(\tilde{v}) &
   \,\,\lower 2pt \hbox{$\buildrel<\over{\scriptstyle{\sim}}$}\,\, -10.8
\end{eqnarray}
The difference between the leading logarithm, ln($Q^2/\Lambda^2$), and our
actual result is small for the $1/Q^4$ term (vector correlator),
but larger for the $1/Q^6$ terms in the various correlators.  In the case of
the scalar correlator the  difference is remarkably enormous.
The leading logarithm still controls the
$Q^2$ dependence, but the magnitude of the implied condensates at a
given value of $Q^2$ has been shifted substantially.  We discuss the
implications in the conclusion.

\vspace{5mm}
{\flushleft\large\bf The {\boldmath $\pi^+ - \pi^0$} electromagnetic
                     mass difference}
\vspace{5mm}

Of most physical interest
is the vector minus axialvector, or left-right (LR) correlator, which is a
direct measure of chiral symmetry breaking.  For $\cal M$=0,
\begin{equation}
   {}^{LR}\tilde\Pi(Q^2)\delta^{ab} = {}^V\tilde\Pi^{ab}(Q^2) -
                                      {}^A\tilde\Pi^{ab}(Q^2).
\end{equation}
We now return to our full expression for the LR correlator, without a
$1/Q^2$ expansion, and use it to calculate the $\pi^+ - \pi^0$ electromagnetic
mass difference.  This mass difference is given by\cite{Das}
\begin{equation}\label{massdiff}
   {\Delta}m_\pi^2 = \frac{3\alpha}{4{\pi}f_\pi^2}\int_0^\infty
   dQ^2~\left[Q^2~{}^{LR}\tilde\Pi(Q^2)\right]
\end{equation}
where $\alpha$ is the electromagnetic coupling.
We will neglect the tiny effects of nonzero current quark masses.  As discussed
earlier, the large(small) $Q^2$ portion of the correlator will be obtained
from Fig.~\ref{fig:loop}(Fig.~\ref{fig:GNC} plus meson loops).
In fact, it will become evident
as we proceed that we can get an upper bound on the mass difference by simply
using Fig.~\ref{fig:loop} for all $Q^2$.

We must also decide how to extend the asymptotic form of $\Sigma(Q^2)$ given in
(\ref{massform}) to smaller $Q^2$.  The following simple ansatz
contains four parameters: $Am_0^2$, $Bm_0^3$, $C$ and $M^2$.
(As in \cite{GNC}, $m_0$ represents the scale of the constituent quark mass.
Since it is not an independent parameter here, we are free to choose it to be
numerically identical to its value in \cite{GNC}.)
\begin{equation}\label{fullmass}
   \Sigma(Q^2) = \frac{Bm_0^3}{\left[Am_0^2+Q^2\right]\left[1+C\,\left\{
                 {\rm ln}\left(1+\frac{Q^2}{M^2}\right)\right\}^{1-d}\right]}
\end{equation}
The motivation for this functional form comes from the $M \rightarrow \infty$
limit, where $\Sigma(Q^2)$ becomes the mass function used originally in the
low energy GNC model.  We wish to include the correct logarithmic
behavior of the mass function at high energies without making significant
changes to the original low energy form.  The ten dimensionless quantities
$L_i(\mu)$ which appear in the standard chiral Lagrangian\cite{Gasser} have
been obtained from the GNC model (i.e. Fig.~\ref{fig:GNC})
in the $M \rightarrow \infty$ limit.\cite{GNC}\cite{moreGNC}
They are sensitive to the
shape of the mass function, but not to the overall scale, so from this we can
determine the value of $Am_0^2$.
\begin{equation}
   (470~{\rm MeV})^2
   \,\,\lower 2pt \hbox{$\buildrel<\over{\scriptstyle{\sim}}$}\,\, Am_0^2
   \,\,\lower 2pt \hbox{$\buildrel<\over{\scriptstyle{\sim}}$}\,\,
   (550~{\rm MeV})^2
\end{equation}

One constraint on the three remaining parameters in (\ref{fullmass})
comes from demanding that $\Sigma(Q^2)$ satisfies the high energy behavior
of (\ref{massform}), where the numerical value of
$\left<{\overline q}q\right>_\mu$ is known
phenomenologically.  This gives
\begin{equation}\label{constraint1}
   \frac{Bm_0^3}{C} \approx \frac{-4d\pi^2}{3}\left<{\overline q}q\right>_\mu
   \left({\rm ln}\frac{\mu^2}{\Lambda^2}\right)^{-d}
\end{equation}
Two more constraints arise from the low energy behavior
of the LR correlator, which must reproduce the experimentally determined
coefficients of the chiral Lagrangian.  In the notation of \cite{Gasser},
the relevant coefficients are $F_0$ and $L_{10}$.
\begin{equation}\label{lowenergy}
   Q^2~{}^{LR}\tilde\Pi(Q^2) = F_0^2 + 4Q^2L_{10}(\mu) -
   \frac{N_fQ^2}{6(4\pi)^2}\left[\frac{5}{3}-{\rm ln}\left(\frac{Q^2}{\mu^2}
   \right)\right] + {\cal O}(Q^4)
\end{equation}
$N_f$ is the number of quark flavors.
The dependence of $L_{10}$ on the renormalization scale $\mu$ is canceled by
the logarithm (which comes from internal meson loops),
so that ${}^{LR}\tilde\Pi(Q^2)$ is independent of $\mu$.  We will eliminate
the parameters $Bm_0^3$ and $C$ by using the phenomenological values for
$\left<{\overline q}q\right>_\mu$ and $F_0$.
The final parameter $M$ will be determined by requiring the mass of
(\ref{fullmass}) in the calculation of Fig.~\ref{fig:GNC} to produce a value
for
$L_{10}$ which is within, say, 25\% of the original GNC result.

Before proceeding, we point out an interesting relation between the high
and low energy calculations.
{}From the evaluation of Fig.~\ref{fig:GNC}, we obtain
\begin{equation}\label{fpi}
   F_0^2 = \frac{3}{8\pi^2}\int_0^{\infty}
             ds\frac{s{\Sigma}(s)\left[2{\Sigma}(s)
             -s\Sigma^{\prime}(s)\right]}{\left[s+\Sigma^2(s)\right]^2}
\end{equation}
and a more lengthy expression for $L_{10}$.  It turns out that
exactly the same expression for $F_0$ happens to come from
Fig.~\ref{fig:loop}.
The two expressions for $L_{10}$ are not the same;
Fig.~\ref{fig:loop} correctly predicts that $L_{10}$ is negative, but
the magnitude is only about half of the correct GNC result.  This
means that if Fig.~\ref{fig:loop} is used for all momenta,
the slope of the LR correlator at $Q^2$=0 is too shallow
(see Fig.~\ref{fig:VVAA}), and an upper bound
is obtained for ${\Delta}m_\pi^2$.  We choose the strongest upper bound by
using the smallest value of $M$ which keeps $L_{10}$ (of Fig.~\ref{fig:GNC})
within 25\% of the original ($M \rightarrow \infty$) GNC result.

For definiteness, we use $d$=4/9 with the following inputs,
\begin{eqnarray}\label{expt1}
   F_0 & \approx & 88{\rm MeV} \label{expt2} \\
   \left<{\overline q}q\right>_\mu\left({\rm ln}\,\frac{\mu^2}
   {\Lambda^2}\right)^{-d} &
   \approx & -(220{\rm MeV})^3
\end{eqnarray}
Note that this last expression is independent of $\mu$.
With these values, the parameters $B$ and $C$ of (\ref{fullmass}) are of
order unity, and the scale $M$ is about 3 GeV.
Fig.~\ref{fig:loop} then gives an upper bound on
the electromagnetic pion mass difference.
\begin{equation}
   m_{\pi^+}-m_{\pi^0}
   ,\,\lower 2pt \hbox{$\buildrel<\over{\scriptstyle{\sim}}$}\,\, 5.1~{\rm MeV}
\end{equation}

To obtain a direct estimate we will calculate the high energy expression
for the LR correlator from Fig.~\ref{fig:loop} down to some intermediate
scale $Q^2_{high}$ and use Fig.~\ref{fig:GNC} plus meson loops
below the scale $Q^2_{low}$ such that
\begin{equation}\label{Rdef}
   Q^2~{}^{LR}\tilde\Pi(Q^2_{low}) = Q^2~{}^{LR}\tilde\Pi(Q^2_{high})
   \equiv R^2
\end{equation}
Between $Q^2_{low}$ and $Q^2_{high}$, this function will be approximated
by the constant $R^2$.  The LR correlator is plotted versus $Q^2$ in
Fig.~\ref{fig:VVAA}, and our result for the pion mass difference is
shown in Fig.~\ref{fig:mdiff} as a function of $R$.
The experimental value (after subtracting $m_d-m_u$ effects) is\cite{Gasser}
\begin{equation}
   \left[m_{\pi^+}-m_{\pi^0}\right]_{expt} = 4.43 \pm 0.03~{\rm MeV}
\end{equation}
and corresponds to $50{\rm MeV}
\,\,\lower 2pt \hbox{$\buildrel<\over{\scriptstyle{\sim}}$}\,\,
R \,\,\lower 2pt \hbox{$\buildrel<\over{\scriptstyle{\sim}}$}\,\, 55{\rm MeV}$.
 This implies
that for this calculation, our low energy model is good up to a scale of order
$Q^2_{low} \sim 400-450{\rm MeV}$ and our high energy model is good down to a
scale of order $Q^2_{high} \sim 750-850{\rm MeV}$.  Both of these scales are
very reasonable.

\vspace{5mm}
{\flushleft\large\bf Conclusions}
\vspace{5mm}

One consequence of the dynamical breakdown of chiral symmetry in QCD is the
generation of a momentum dependent light quark mass.  In the context of the
vector, axialvector and scalar two-point correlators we have shown how
this effective mass can be included systematically in calculations.
The Ward-Takahashi identities and chiral symmetry are respected, as are the
first few terms (at least) of the operator product expansion (OPE).
An interesting result is the existence of terms at order $1/Q^6$ which do not
contain the leading logarithm but which are not insignificant, especially
in the case of the scalar correlator.

It appears that for the scalar two-point function a
naive application of the OPE in conjunction with the
vacuum saturation approximation does not adequately describe some
expected physics of QCD, namely the physics associated with the dynamical quark
mass.  This is perhaps not surprising.  Practitioners of sum rules have
long claimed\cite{Nov} that there is something deficient in the usual
application of the OPE in the case of the scalar and pseudoscalar two-point
functions.  This is particularly clear in the pseudoscalar case, for which the
conventional OPE does not adequately account for the pion appearing in the
sum rule.  This has led to speculations of additional contributions to these
OPEs, including instantons\cite{Nov}, renormalons\cite{VZ}, and effective
four-fermion interactions\cite{Yam}.  Our work indicates that the additional
contributions will also have to reflect effects associated with
chiral symmetry breaking, and in particular the momentum dependence of the
dynamical quark mass.  On the other hand in the case of the vector and
axialvector two-point functions, the subleading terms in our analysis are
not substantial enough to say that there is a serious conflict with standard
treatments.

We have described in this work a minimal model.  It could be
extended for example by including an effective wavefunction
renormalization parameter in the quark propagator, or by adding extra terms
to the vertices which maintain the Ward-Takahashi identities and the OPE.
Interestingly enough, we find that the minimal model is sufficient to
account for the $\pi^+ - \pi^0$ electromagnetic mass difference.

\vspace{5mm}
{\flushleft\large\bf Acknowledgements}
\vspace{5mm}

We thank Michael Luke for useful discussions.
This research was supported in part by the Natural Sciences and Engineering
Research Council of Canada.



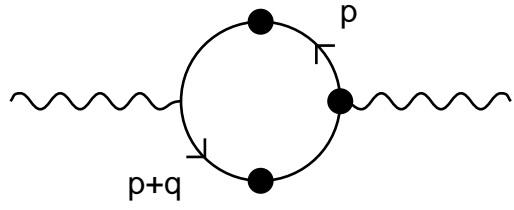
\begin{figure}[p]
\caption{The complete QCD contribution to a two-point
         correlator.  Both propagators are full propagators; one vertex is
         full and the other vertex is a bare $\frac{\lambda^a}{2}[\gamma_5]$ or
         $\frac{\lambda^a}{2}\gamma_\mu[\gamma_5]$.}\label{fig:loop}
\end{figure}
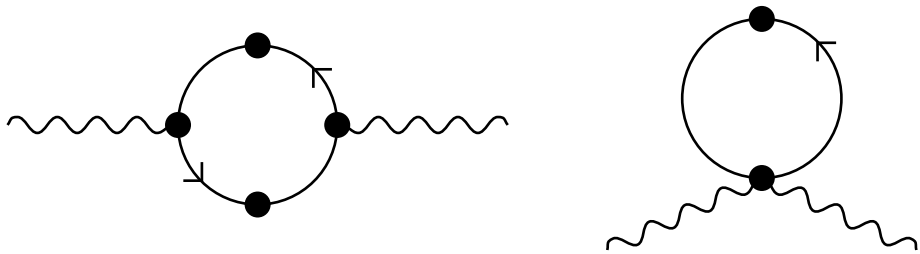
\begin{figure}[p]
\caption{The quark contribution to a two-point correlator in the
         GNC$^\prime$ model.  Meson contributions are not shown.
         }\label{fig:GNC}
\end{figure}
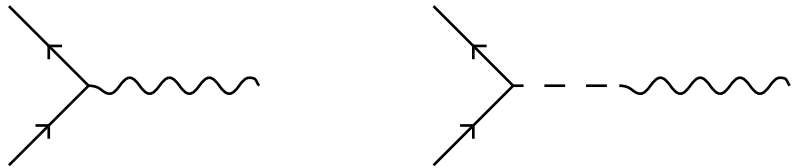
\begin{figure}[p]
\caption{The two distinct components of the full axialvector and pseudoscalar
         vertices as derived from the GNC$^\prime$ model.
         The dashed line represents a pseudoscalar meson propagator which is
         generated from the diagrams of Fig.~\protect\ref{fig:prop}.
         }\label{fig:pole}
\end{figure}
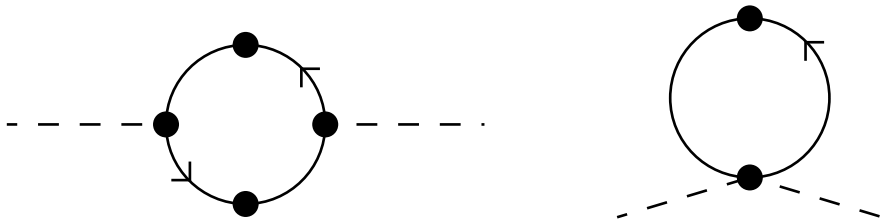
\begin{figure}[p]
\caption{The GNC$^\prime$ Lagrangian does not contain an explicit meson
         propagator, but a propagator is generated by the quark loop diagrams
         shown here.}\label{fig:prop}
\end{figure}
\begin{figure}[p]
\caption{The vector minus axialvector two-point correlator obtained from the
         GNC$^\prime$ model at low energies and from
         Fig.~\protect\ref{fig:loop} at high energies for a typical choice
         of parameters.  The two pieces are
         matched to a constant, $R^2$, in the intermediate region.
         }\label{fig:VVAA}
\end{figure}
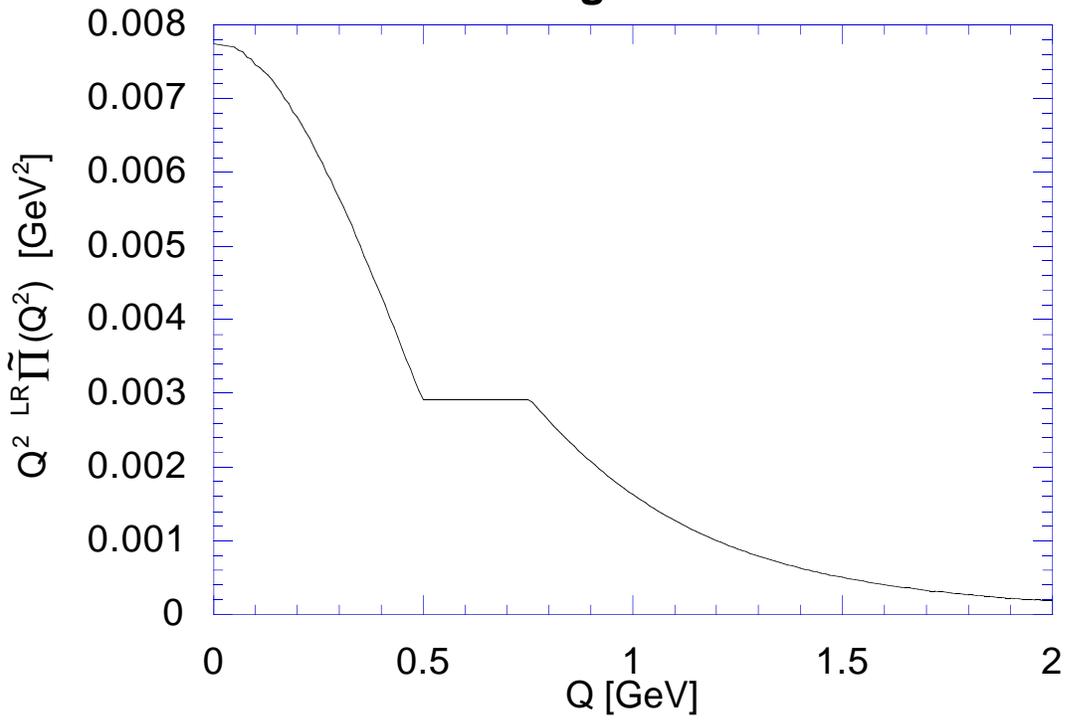
\begin{figure}[p]
\caption{The electromagnetic mass difference of the pion as a function of $R$,
         defined by (\protect\ref{Rdef}), for a typical choice of parameters.
         }\label{fig:mdiff}
\end{figure}

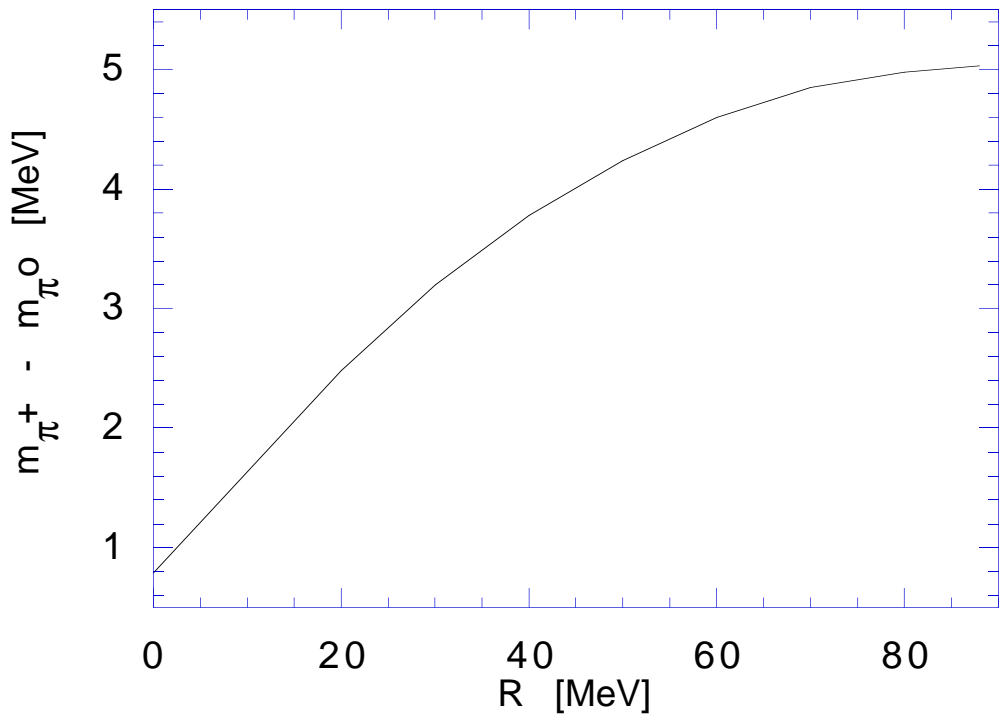
\end{document}